\documentclass[12pt]{article}
\usepackage{amsmath}
\usepackage{amssymb}

\input{tcilatex}

\begin{document}

\bigskip

\bigskip \ 

\bigskip \ 

\begin{center}
\textbf{DUALITY, MATROIDS, QUBITS, TWISTORS}

\smallskip \ 

\textbf{AND\ SURREAL NUMBERS}

\smallskip \ 

\smallskip \ 

\smallskip \ 

J. A. Nieto\footnote{%
nieto@uas.edu.mx, janieto1@asu.edu}

\smallskip \ 

\textit{Facultad de Ciencias F\'{\i}sico-Matem\'{a}ticas de la Universidad
Aut\'{o}noma de Sinaloa, 80010, Culiac\'{a}n Sinaloa, M\'{e}xico}

\bigskip \ 

\bigskip \ 

Abstract
\end{center}

We show that \textit{via} the Grassmann-Pl\"{u}cker relations, the various
apparent unrelated concepts, such as duality, matroids, qubits, twistors and
surreal numbers are, in fact, deeply connected. Moreover, we conjecture the
possibility that these concepts may be considered as underlying mathematical
structures in quantum gravity.

\bigskip \ 

\bigskip \ 

\bigskip \ 

\bigskip \ 

\bigskip \ 

\bigskip \ 

\bigskip \ 


Pacs numbers: 04.60.-m, 04.65.+e, 11.15.-q, 11.30.Ly

September, 2018

\newpage

It is a fact that the duality concept is everywhere in both mathematics and
physics. Of course, since the list of examples of this fact is very large
and since we are concern with quantum gravity let us just briefly mention,
as examples in which the duality concept plays a fundamental role, matroid
theory [1]-[2] (see also Refs. [3]-[9] and references therein) and surreal
numbers [10]-[12] in mathematics and string theory [13] and loop quantum
gravity [14] in physics. The origin of matroid theory can be traced back to
graph theory were according to the Kuratowski theorem a graph has a dual if
does not contain the complete graphs $K_{5}$ and $K_{3,3}$ (see Ref. [15]).
A matroid is a generalization of the graph concept in which every matroid
has a dual. One may understand why matroid theory is a generalization of
graph theory by associating with every graph $G$ a matroid $M(G)$. So one
must have $M(K_{5})$ and $M(K_{3,3})$, but according to matroid theory one
must have the corresponding duals $M^{\ast }(K_{5})$ and $M^{\ast }(K_{3,3})$
which turns out to be non-graphic. A surreal number $x=\{X_{L},X_{R}\}$ is
written in terms of the dual sets $X_{L}$ left set and $X_{R}$ the right set
which satisfies two main axioms (see below). Surprisingly these dual numbers
contains the structure of real numbers among other numerical structures. On
the other hand it is known that the origin of $M$-theory [16] was inspired
by trying to make sense of a number of dualities between string theory and $%
p $-branes. For instance, in eleven dimensions the $1$-brane is dual to the $%
5$-brane (see Ref. [16]). Finally, it is known that loop quantum gravity
emerges from the discovery of the Ashtekar variables which in turn arises by
the requirement of the canonical formalism applied to the self-dual Ricci
curvature tensor (see [14] and references therein).

Of course, the duality concepts refereed above may be at first sight quite
different for each example. So the first step it is to introduce a formal
definition of the concept of duality. It turns out that at least in matroid
theory one finds such a formal definition [17]. Let $\mathcal{M}$ denote the
family of all matroids $M$ which corresponding to the ground set $E$. The
matroid duality is a map $^{\ast }:\mathcal{M}\rightarrow \mathcal{M}$
satisfying the two main axioms:

\smallskip \ 

(a) $^{\ast }{}^{\ast }M=M$ \ $\  \  \  \  \  \  \  \  \  \  \  \ (\forall M\in 
\mathcal{M}).$

(b) $E(^{\ast }M)=E(M)\  \  \  \  \  \ (\forall M\in \mathcal{M}).$

\smallskip \ 

Inspired by this definition of duality in oriented matroid theory let us
propose a general tensor definition of duality structure. Consider a family$%
\mathcal{A}$ of all completely antisymmetric tensors $A$ ($p$-forms), which
correspond to space of dimension $d$, together with an operation $+$ which
can be any well defined tensorial sum operation. The pair $(\mathcal{A},+)$
determines a dual structure through the map $^{\ast }:\mathcal{A}\rightarrow 
\mathcal{A\ }$if satisfies the following axioms:

\smallskip \ 

(I) $^{\ast }{}^{\ast }A=A\  \  \  \  \  \  \  \  \  \  \  \ (\forall A\in \mathcal{A}%
). $

(II) $d(^{\ast }A)=d(A)$\  \  \  \  \ $(\forall A\in \mathcal{A}).$

\smallskip \ 

Note that (II) plays the role of (b) in matroid theory.

\smallskip \ 

Assuming the particular case that $\mathcal{A}$ corresponds to family of
zero-rank tensors one may add two additional axioms, namely

\smallskip \ 

(III) There exist in $\mathcal{A}$ a self dual element $^{\ast }0=0$ such
that $A+0=0+A=A$, $\  \ (\forall A\in \mathcal{A)}$.

(IV) For $\forall A\in \mathcal{A}$ one has $A+$ $^{\ast }A=$ $^{\ast }A+A=0$%
.

\smallskip \ 

One can prove that the element $0$ in (III) is unique as follows: Assume
that $(\mathcal{A},+)$ is a dual structure with two self-dual elements $0$
and $0^{\prime }$. Then $0=0+0^{\prime }=0^{\prime }$. Moreover, according
to the axiom (IV) the element $^{\ast }A$ can be considered as the inverse
of $A$. In order to show that the inverse $^{\ast }A$ is unique one takes
recourse of the axiom (I) instead of the associativity axiom in group
theory. In fact, assume that an arbitrary element $A$ in $\mathcal{A}$ has
two inverses $^{\ast }A$ and $^{\ast }B$. Thus, one has (i) $A+$ $^{\ast
}A=0 $ and (ii) $A+$ $^{\ast }B=0$. Applying the axioms (I) and (III) to
(ii) one obtains $^{\ast }A+$ $B=0$ and therefore according (i) one gets $%
^{\ast }A+$ $B=$ $^{\ast }A+A$ which means that $B=$ $A$. The two axioms
(III) and (IV) are similar to the definition of a field in number theory.
For these reasons one it is straightforward to verify that the integer $Z$
and the real numbers $R$ are in fact dual structures.

The main goal of the present work is to comment about the possibility that
the various concepts such as oriented matroids, qubits, twistors and surreal
numbers are linked by the duality symmetry. Moreover we shall argue that
such a dual concept may be considered as an underlying mathematical tool in
quantum gravity.

It turns out that the completely antisymmetric $\varepsilon $-symbol becomes
the underlying mathematical object in all these connections. Specifically,
the $\varepsilon $-symbol can be defined as%
\begin{equation}
\varepsilon ^{a_{1}...a_{d}}\in \{-1,0,1\}.  \label{1}
\end{equation}%
Here, the indices $a_{1},...,a_{d}$ run from $1$ to $d$. This is a $d$-rank
density tensor which values are $+1$ or $-1$ depending on even or odd
permutations of $\varepsilon ^{12...d}$, respectively. Moreover, $%
\varepsilon ^{a_{1}...a_{d}}$ takes the value $0$ unless the values of $%
a_{1}...a_{d}$ are all different. Lowering and rising the indices with a
Kronecker delta $\delta _{ab}$ one finds that

\begin{equation}
\varepsilon ^{a_{1}...a_{d}}\varepsilon _{b_{1}...b_{d}}=\delta
_{b_{1}...b_{d}}^{a_{1}...a_{d}},  \label{2}
\end{equation}%
where $\delta _{b_{1}...b_{d}}^{a_{1}...a_{d}}$ is a generalized Kronecker
delta. A contraction in (2) of the last $n$-indices of the type $a_{i}$ with
the last $n$-indices of the type $b_{i}$ leads to

\begin{equation}
\varepsilon ^{a_{1}...a_{k-1}a_{k}...a_{d}}\varepsilon
_{b_{1}...b_{k-1}a_{k}...a_{d}}=n!\delta
_{b_{1}...b_{k-1}}^{a_{1}...a_{k-1}},  \label{3}
\end{equation}%
with $n=d-k+1$. In particular one has

\begin{equation}
\varepsilon ^{a_{1}...a_{d}}\varepsilon _{a_{1}...a_{d}}=d!.  \label{4}
\end{equation}

Let $v_{a}^{i}$ be any $d\times n$ matrix over some field $F$, where the
index $i$ takes values in the ordinal set $E=\{1,...,n\}$. Consider the
object%
\begin{equation}
\Sigma ^{i_{1}...i_{d}}=\varepsilon
^{a_{1}...a_{d}}v_{a_{1}}^{i_{1}}...v_{a_{d}}^{i_{d}}.  \label{5}
\end{equation}%
Using the $\varepsilon $-symbol property

\begin{equation}
\varepsilon ^{a_{1}...[a_{d}}\varepsilon ^{b_{1}...b_{d}]}=0,  \label{6}
\end{equation}%
it is not difficult to prove that $\Sigma ^{i_{1}...i_{d}}$ satisfies the
Grassmann-Pl\"{u}cker relations (see [18] and references therein), namely

\begin{equation}
\Sigma ^{i_{1}...[i_{d}}\Sigma ^{j_{1}...j_{d}]}=0.  \label{7}
\end{equation}%
Here, the brackets in the indices of (6) and (7) mean completely
antisymmetric.

Through (5) one can define the object%
\begin{equation}
\mathbf{\Sigma }=\frac{1}{d!}\Sigma ^{i_{1}...i_{d}}e_{i_{1}}\wedge
e_{i_{2}}\wedge ...\wedge e_{i_{d}},  \label{8}
\end{equation}%
where $e_{i_{1}},e_{i_{2}},...,e_{i_{d}}$ are $1$-form bases associated \
with the $(_{d}^{n})$-dimensional real vector space of alternating $d$-forms
on $R^{n}$. It turns out that (8) can also be written as

\begin{equation}
\mathbf{\Sigma }=\mathbf{v}_{1}\wedge \mathbf{v}_{2}\wedge ...\wedge \mathbf{%
v}_{d},  \label{9}
\end{equation}%
for some $\mathbf{v}_{1},\mathbf{v}_{2},...,.\mathbf{v}_{d}\in R^{n}$. This
means that $\Sigma ^{i_{1}...i_{d}}$ corresponds to an alternating
decomposable $d$-form [19].

A realizable chirotope $\chi $ is defined as

\begin{equation}
\chi ^{i_{1}...i_{d}}=sign\Sigma ^{i_{1}...i_{d}}.  \label{10}
\end{equation}%
In order to define non-realizable chirotopes it is convenient to write the
expression (7) in the alternative form

\begin{equation}
\tsum \limits_{k=1}^{d+1}s_{k}=0,  \label{11}
\end{equation}%
where

\begin{equation}
s_{k}=(-1)^{k}\Sigma ^{i_{1}...i_{d-1}j_{k}}\Sigma ^{j_{1}...\hat{\jmath}%
_{k}...j_{d+1}}.  \label{12}
\end{equation}%
Here, $j_{d+1}=i_{d}$ and $\hat{\jmath}_{k}$ establish the notation for
omitting this index. Thus, for a general definition, one defines a $d$-rank
chirotope $\chi :E^{d}\rightarrow \{-1,0,1\}$ if there exist $%
r_{1},...,r_{d+1}\in R^{+}$ such that

\begin{equation}
\sum \limits_{k=1}^{d+1}r_{k}s_{k}=0,  \label{13}
\end{equation}%
with

\begin{equation}
s_{k}=(-1)^{k}\chi ^{i_{1}...i_{d-1}j_{k}}\chi ^{j_{1}...\hat{\jmath}%
_{k}...j_{d+1}},  \label{14}
\end{equation}%
and $k=1,...,d+1$. It is evident that (11) is a particular case of (13).
Therefore, there are chirotopes that may be non-realizable. Moreover, this
definition of a chirotope admits a straightforward generalization to the
complex structure setting. In this case the complex chirotopes are called
phirotopes [20]-[22].

Given a chirotope (or phirotope) $\chi ^{i_{1}...i_{d}}$ its dual is defined
as

\begin{equation}
^{\ast }\chi _{i_{d+1}...i_{p}=}\varepsilon
_{i_{1}...i_{d}i_{d+1}...i_{p}}\chi ^{i_{1}...i_{d}}.  \label{15}
\end{equation}%
Here $D=d+p$ is the total dimension of the ground state $E$. Observe that
due to the relations (3) one gets 
\begin{equation}
^{\ast \ast }\chi =\chi ,  \label{16}
\end{equation}%
which means that $\chi $ satisfies the axiom (I). It turns out that (16) is
true for a general completely antisymmetric object $F$ ($d$-form) when its
dual is defined in terms of the $\varepsilon $-symbol. In fact, when $D$ is
even one can write $D=d+d=2d$ and in this case one can define the self-dual
(antiself-dual) tensor as

\begin{equation}
^{\pm }F=F\pm ^{\ast }F  \label{17}
\end{equation}%
One observe that $^{\pm }F$ satisfies

\begin{equation}
^{\ast \pm }F=\pm ^{\pm }F  \label{18}
\end{equation}%
Thus, one sees that for $D$ even the $^{\pm }F$ tensor is another self-dual
(antiself-dual) notion other than the $0$ element in the axioms (III) and
(IV).

Let us now explain how the Grassmann-Pl\"{u}cker relation (7) is connected
with qubit theory (see Ref. [23] and references therein). For this purpose
consider the general complex state $\mid \psi >\in C^{2^{N}}$

\begin{equation}
\mid \psi >=\dsum
\limits_{A_{1},A_{2},...,A_{N}=0}^{1}Q_{A_{1}A_{2}...A_{N}}\mid
A_{1}A_{2}...A_{N}>,  \label{19}
\end{equation}%
where the states $\mid A_{1}A_{2}...A_{N}>=\mid A_{1}>\otimes \mid
A_{2}>...\otimes \mid A_{N}>$ correspond to a standard basis of the $N$%
-qubit. For a $3$-qubit (19) becomes

\begin{equation}
\mid \psi >=\dsum \limits_{A_{1},A_{2},A_{3}=0}^{1}Q_{A_{1}A_{2}A_{3}}\mid
A_{1}A_{2}A_{3}>,  \label{20}
\end{equation}%
while for $4$-qubit one has

\begin{equation}
\mid \psi >=\dsum
\limits_{A_{1},A_{2},A_{3},A_{4}=0}^{1}Q_{A_{1}A_{2}A_{3}A_{4}}\mid
A_{1}A_{2}A_{3}A_{4}>.  \label{21}
\end{equation}%
It turns out that, in a particular subclass of $N$-qubit entanglement, the
Hilbert space can be broken into the form $C^{2^{N}}=C^{L}\otimes C^{l}$,
with $L=2^{N-n}$ and $l=2^{n}$. Such a partition allows a geometric
interpretation in terms of the complex Grassmannian variety $Gr(L,l)$ of $l$%
-planes in $C^{L}$ via the Pl\"{u}cker embedding. It turns out that in this
scenario the complex $3$-qubit, $4$-qubit admit a geometric interpretation
in terms of the complex Grassmannian varieties $Gr(4,2)$, $Gr(8,2)$,
respectively (see Ref. [23] for details). The idea is to associate the first 
$N-n$ and the last $n$ indices of $Q_{A_{1}A_{2}...A_{N}}$ with a $L\times l$
matrix $\omega _{a_{1}}^{i_{1}}$. This can be interpreted as the coordinates
of the Grassmannian $Gr(L,l)$ of $l$-planes in $C^{L}$. Using the matrix $%
\omega _{p_{1}}^{i_{1}}$ one can define the Pl\"{u}cker coordinates

\begin{equation}
\mathcal{Q}^{i_{1}...i_{d}}=\varepsilon ^{a_{1}...a_{d}}\omega
_{a_{1}}^{i_{1}}...\omega _{a_{d}}^{i_{d}},  \label{22}
\end{equation}%
which one recognizes as the complex version of the decomposable tensor $%
\Sigma ^{i_{1}...i_{d}}$ defined in (5). Moreover, one verifies that under
the transformation $\omega \rightarrow S\omega $ with $S\in GL(l,C)$ the Pl%
\"{u}cker coordinates transform as $\mathcal{Q}^{i_{1}...i_{d}}\rightarrow
Det(S)\mathcal{Q}^{i_{1}...i_{d}}$ and of course $\mathcal{\Sigma }%
^{i_{1}...i_{d}}$ satisfies the Grassmann-Pl\"{u}cker relations

\begin{equation}
\mathcal{Q}^{i_{1}...[i_{d}}\mathcal{Q}^{j_{1}...j_{d}]}=0.  \label{23}
\end{equation}

Now, consider the quantity $\sigma _{\mu }=(\sigma _{0},\sigma _{\hat{\imath}%
})$, where the $\sigma _{\hat{\imath}}$ denotes Pauli matrices and $\sigma
_{0}$ is the identity matrix. By using $\sigma _{\mu }$ the linear momentum
in $4$-dimensions $p^{\mu }$ can be written as

\begin{equation}
p^{A\dot{B}}=\sigma _{\mu }^{A\dot{B}}p^{\mu }.  \label{24}
\end{equation}%
This is the spinorial representation of $p^{\mu }$. An interesting aspect
emerges if one sets $Det(p^{A\dot{B}})=0$, corresponding to a null momentum $%
p^{\mu }p_{\mu }=0$. This means that

\begin{equation}
\frac{1}{2!}\varepsilon _{AC}\varepsilon _{\dot{B}\dot{D}}p^{A\dot{B}}p^{C%
\dot{D}}=0.  \label{25}
\end{equation}%
A solution to this equation is given by $p^{A\dot{B}}=\xi ^{A}\eta ^{\dot{B}%
} $. Since $p^{\mu }$ is real vector one verifies that $p^{A\dot{B}}=\bar{p}%
^{\dot{B}A}$ and therefore 
\begin{equation}
\xi ^{A}\eta ^{\dot{B}}=\bar{\xi}^{\dot{B}}\bar{\eta}^{A}.  \label{26}
\end{equation}%
One finds that this last expression means that $\eta ^{\dot{B}}=a\bar{\xi}^{%
\dot{B}}$, where due to (26) one sees that $a$ is real number which can be
normalized in the form $a=\pm $. So one has found that, in the case of null
momentum, one can write $p^{A\dot{B}}$ in terms of a more fundamental
complex quantity $\xi ^{A}$, namely%
\begin{equation}
p^{A\dot{B}}=\pm \xi ^{A}\bar{\xi}^{\dot{B}}.  \label{27}
\end{equation}%
Similar analysis applies to the total angular momentum $M^{\mu \nu }=x^{\mu
}p^{\nu }-x^{\nu }p^{\mu }+S^{\mu \nu }$, where $S^{\mu \nu }$ is the
internal angular momentum satisfying the Tulczyjew second class constraint
[24];

\begin{equation}
S^{\mu \nu }p_{\nu }=0.  \label{28}
\end{equation}%
Observe that due to (28) and since $p^{\mu }$ is a null vector one has $%
M^{\mu \nu }p_{\nu }=-(x^{\nu }p_{\nu })p^{\mu }$ This means that $\delta
_{\alpha \beta \gamma }^{\tau \mu \nu }p^{\alpha }M^{\beta \gamma }p_{\nu
}=0 $. In turn this leads to $\varepsilon _{\sigma \alpha \beta \gamma
}\varepsilon ^{\sigma \tau \mu \nu }p^{\alpha }M^{\beta \gamma }p_{\nu }=0$
or $\varepsilon _{\sigma \alpha \beta \gamma }\varepsilon ^{\sigma \tau \mu
\nu }p^{\alpha }S^{\beta \gamma }p_{\nu }=0$. Therefore, if one defines the $%
4$-vector $S_{\sigma }=\frac{1}{2}\varepsilon _{\sigma \alpha \beta \gamma
}p^{\alpha }S^{\beta \gamma }$ one obtains $\varepsilon ^{\sigma \tau \mu
\nu }S_{\sigma }p_{\nu }=0$ and consequently one discovers that

\begin{equation}
S_{\mu }=sp_{\mu ,}  \label{29}
\end{equation}%
for some non-vanishing constant $s$ which is identified with the helicity of
the system. This means that the spin $S_{\mu }$ is parallel or anti-parallel
to $p_{\mu }$ depending of the sign of $s$. So, determining $p^{A\dot{B}}$
in terms of $\xi ^{A}$ via (27) is equivalent to determine $S^{A\dot{B}}$ in
the form $S^{A\dot{B}}=s\xi ^{A}\bar{\xi}^{\dot{B}}$. Thus, considering (28)
one sees that the left relevant part of $M^{\mu \nu }$ is 
\begin{equation}
L^{\mu \nu }=x^{\mu }p^{\nu }-x^{\nu }p^{\mu },  \label{30}
\end{equation}%
but again since $p^{\mu }$ is a null vector one has $L^{\mu \nu }p_{\nu
}=-(x^{\nu }p_{\nu })p^{\mu }$ which means that out of the six true degrees
of freedom of $L^{\mu \nu }=-L^{\nu \mu }$ three are already determined by $%
p^{\mu }$. Therefore, the corresponding spinor representation of $L^{\mu \nu
}$, namely $L^{A\dot{B}C\dot{D}}=\sigma _{\mu }^{A\dot{B}}\sigma _{\nu }^{C%
\dot{D}}L^{\mu \nu }$, can be written as

\begin{equation}
L^{A\dot{B}C\dot{D}}=\mu ^{AC}\epsilon ^{\dot{B}\dot{D}}+\epsilon ^{AC}\mu ^{%
\dot{B}\dot{D}}.  \label{31}
\end{equation}%
Here, $\mu ^{AC}=\mu ^{CA}$ is a symmetric matrix and therefore has only
three independent complex degrees of freedom. In order to reduce $\mu ^{AC}$
to only three real components which of course are related to the true three
degrees of freedom of $L^{\mu \nu }$ one further writes $\mu ^{AC}$ in the
form $\mu ^{AC}=\xi ^{A}\pi ^{C}+\xi ^{C}\pi ^{A}$. If to the coordinates $%
\xi _{\dot{A}}$ one adds the the spinor $\pi ^{A}$ one is lead to the
twistor structure $\mathcal{P}^{\alpha }=(\pi ^{A},\xi _{\dot{A}})$ [25]
(see Ref. [26] and references therein) which can be identified with a point
in $C^{4}$. This analysis revel that in the case of a null system the eight
coordinates $(x^{\mu },p^{\nu })$ in $R^{8}$ may in principle be associated
with the coordinates $(\pi ^{A},\xi _{\dot{A}})$ in the twistor complex
space $C^{4}$.

Consider the $2$-index twistor

\begin{equation}
P^{\alpha \beta }=\mathcal{P}_{1}^{\alpha }\mathcal{P}_{2}^{\beta }-\mathcal{%
P}_{2}^{\alpha }\mathcal{P}_{1}^{\beta },  \label{32}
\end{equation}%
which can also be written as

\begin{equation}
\mathcal{P}^{\alpha \beta }=\varepsilon ^{ab}\mathcal{P}_{a}^{\alpha }%
\mathcal{P}_{b}^{\beta }.  \label{33}
\end{equation}%
If one defines $p_{1}^{\mu }=x^{\mu }$ and $p_{2}^{\mu }=p^{\mu }$ one sees
that $L^{\mu \nu }$ can be written as

\begin{equation}
L^{\mu \nu }=\varepsilon ^{ab}p_{a}^{\mu }p_{b}^{\nu }  \label{34}
\end{equation}%
and therefore one concludes that $\mathcal{P}^{\alpha \beta }$ can be
understood as the complexification of $L^{\mu \nu }$. Of course, $\mathcal{P}%
^{\alpha \beta }$ satisfies the Grassmann-Pl\"{u}cker relations

\begin{equation}
\mathcal{P}^{\mu \lbrack \nu }\mathcal{P}^{\alpha \beta ]}=0.  \label{35}
\end{equation}%
It turns out that $\mathcal{P}^{\alpha \beta }$ can be used to associate
points in $C^{4}$ with points in the complexified Minkowski space (see Ref.
[25]). From the perspective of oriented complex matroids, $\mathcal{P}%
^{\alpha \beta }$ is just a representable phirotope. One is tempted to
assume that a generalization of twistor theory may be also be associated
with the phirotope theory.

Is it possible that twistors or qubits can be related to surreal number
theory [10]-[12]? Consider the set

\begin{equation}
x=\{X_{L}\mid X_{R}\}  \label{36}
\end{equation}%
and call $X_{L}$ and $X_{R}$ the left and right sets of $x$, respectively.
Conway develops the surreal numbers structure $\mathcal{S}$ from two axioms:

\smallskip \ 

\textbf{Axiom 1}. Every surreal number corresponds to two sets $X_{L}$ and $%
X_{R}$ of previously created numbers, such that no member of the left set $%
x_{L}\in X_{L}$ is greater or equal to any member $x_{R}$ of the right set $%
X_{R}$.

\smallskip \ 

Let us denote by the symbol $\ngeq $ the notion of no greater or equal to.
So the axiom establishes that if $x$ is a surreal number then for each $%
x_{L}\in X_{L}$ and $x_{R}\in X_{R}$ one has $x_{L}\ngeq x_{R}$. This is
denoted by $X_{L}\ngeq X_{R}$.

\smallskip \ 

\textbf{Axiom 2}. One number $x=\{X_{L}\mid X_{R}\}$ is less than or equal
to another number $y=\{Y_{L}\mid Y_{R}\}$ if and only the two conditions $%
X_{L}\ngeq y$ and $x\ngeq Y_{R}$ are satisfied.

\smallskip \ 

This can be simplified by saying that $x\leq y$ if and only if $X_{L}\ngeq y$
and $x\ngeq Y_{R}$.

\smallskip \ 

Observe that Conway definition relies in an inductive method; before a
surreal number $x$ is introduced one needs to know the two sets $X_{L}$ and $%
X_{R}$ of surreal numbers. Using Conway algorithm one finds that at the $j$%
-day one obtains $2^{j+1}-1$ numbers all of which are of form%
\begin{equation}
x=\frac{m}{2^{n}},  \label{37}
\end{equation}%
where $m$ is an integer and $n$ is a natural number, $n>0$. Of course, the
numbers (37) are dyadic rationals which are dense in the reals $R$.

The sum and product of surreal numbers are defined as

\begin{equation}
x+y=\{X_{L}+y,x+Y_{L}\mid X_{R}+y,x+Y_{R}\}  \label{38}
\end{equation}%
and

\begin{equation}
\begin{array}{c}
xy=\{X_{L}y+xY_{L}-X_{L}Y_{L},X_{R}y+xY_{R}-X_{R}Y_{R}\mid X_{L}y+xY_{R} \\ 
\\ 
-X_{L}Y_{R},X_{R}y+xY_{L}-X_{R}Y_{L}\},%
\end{array}
\label{39}
\end{equation}%
respectively. The importance of (38) and (39) is that allow us to prove that
the surreal number structure is algebraically a closed field. Moreover,
through (38) and (39) it is also possible to show that the real numbers $R$
are contained in the surreals $\mathcal{S}$ (see Ref. [10]-[12] for
details). Of course, in some sense the prove relies on the fact that the
dyadic numbers (37) are dense in the reals $R$.

In 1986, Gonshor [12] introduced a different but equivalent definition of
surreal numbers.

\smallskip \ 

\textbf{Definition 1}. A surreal number is a function $f$ from initial
segment of the ordinals into the set $\{+,-\}$.

\smallskip \ 

For instance, if $f$ is the function so that $f(1)=+$, $f(2)=+$, $f(3)=-$, $%
f(4)=+$ then $f$ is the surreal number $(++-+)$. In the Gonshor approach one
obtains the sequence: $1$-day

\begin{equation}
-1=(-),\text{ \  \  \  \  \  \  \  \  \  \  \ }(+)=+1,  \label{40}
\end{equation}%
in the $2$-day

\begin{equation}
-2=(--),\text{ \ }-\frac{1}{2}=(-+),\text{\  \  \ }(+-)=+\frac{1}{2},\text{\  \
\  \  \  \ }(++)=+2,  \label{41}
\end{equation}%
and $3$-day

\begin{equation}
\begin{array}{c}
-3=(---),\text{ \ }-\frac{3}{2}=(--+),\text{\  \ }-\frac{3}{4}=(-+-),\text{\ }%
-\frac{1}{4}=(-++) \\ 
\\ 
(+--)=+\frac{1}{4},\text{ \ }(+-+)=+\frac{3}{4},\text{ \  \ }(++-)=+\frac{3}{2%
},\text{\  \  \  \  \  \ }(+++)=+3,%
\end{array}
\label{42}
\end{equation}%
respectively. Moreover, in Gonshor approach one finds the different numbers
through the formula

\begin{equation}
\mathcal{J}=l\mid \varepsilon _{0}\mid -\frac{\mid \varepsilon _{1}\mid }{2}%
+\sum \limits_{i=2}^{s}\frac{\mid \varepsilon _{i}\mid }{2^{i}},  \label{43}
\end{equation}%
where $\varepsilon _{0},\varepsilon _{1},\varepsilon _{2},...,\varepsilon
_{q}\in \{+,-\}$ and $\varepsilon _{0}\neq \varepsilon _{1}$. Furthermore,
one has $\mid +\mid =+$ and $\mid -\mid =-$. As in the case of Conway
definition, through (43) one gets the dyadic rationals. Just for clarity,
let us consider the additional example: 
\begin{equation}
(++-+-+)=2-\frac{1}{2}+\frac{1}{4}-\frac{1}{8}+\frac{1}{16}=\frac{27}{16}.
\label{44}
\end{equation}%
By defining the order $x<y$ if $x(\alpha )<y(\alpha )$, where $\alpha $ is
the first place where $x$ and $y$ differ and the convention $-<0<+$, it is
possible to show that the Conway and Gonshor definitions of surreal numbers
are equivalent (see Ref. [12] for details).

Suppose that instead of qubits we consider a rebit (real bits) which can be
thought as $j$-tensor [4],

\begin{equation}
t_{A_{1}A_{2}...A_{j}},  \label{45}
\end{equation}%
where the indices $A_{1},A_{2},...,A_{j}$ run from $0$ to $1$. Of course $j$
indicates the rank of $t_{A_{1}A_{2}...A_{j}}$. In tensorial analysis, (45)
is a familiar object. One arrives to a link with surreal numbers by making
the indices identification $0\rightarrow +$ and $1\rightarrow -$. For
instance, the tensor $t_{0010}$ in the Gonshor notation becomes

\begin{equation}
t_{0010}\rightarrow t_{++-+}\rightarrow (++-+).  \label{46}
\end{equation}

In terms of $t_{A_{1}A_{2}...A_{j}}$, the expressions (40), (41) and (42)
read

\begin{equation}
-1=t_{1},\text{ \  \  \  \  \  \  \  \  \  \ }t_{0}=+1,  \label{47}
\end{equation}%
in the $2$-day

\begin{equation}
-2=t_{11,}\text{ \  \ }-\frac{1}{2}=t_{10},\text{\  \  \ }t_{01}=\frac{1}{2},%
\text{\  \  \ }t_{00}=2,  \label{48}
\end{equation}%
and $3$-day

\begin{equation}
\begin{array}{c}
-3=t_{111},\text{ \ }-\frac{3}{2}=t_{110},\text{\  \ }-\frac{3}{4}=t_{101},%
\text{\ }-\frac{1}{4}=t_{100},\text{\ } \\ 
\\ 
t_{011}=+\frac{1}{4},\text{ \ }t_{010}=+\frac{3}{4},\text{ \ }t_{001}=+\frac{%
3}{2},\text{\  \  \ }t_{000}=+3,%
\end{array}
\label{49}
\end{equation}%
\smallskip \ respectively.

Note that there is a duality symmetry between positive and negative labels
in surreal numbers. In fact, one can prove that this is general for any $j$%
-day. This could be anticipated because according to Conway definition a
surreal number can be written in terms of the dual pair left and right sets $%
X_{L}$ and $X_{R}$. Further, the concept of duality it is even clearer in
the Gonshor definition of surreal numbers since in such a case one has a
functions $f$ with the image in the dual set $\{+,-\}$. In terms of the
tensor $t_{A_{1}A_{2}...A_{p}}$ such a duality can be written in the form

\begin{equation}
t_{A_{1}A_{2}...A_{p}}+(-1)^{p}\varepsilon _{A_{1}B_{1}}\varepsilon
_{A_{2}B_{2}}...\varepsilon _{A_{p}B_{p}}t^{B_{1}B_{2}...B_{p}}=0,
\label{50}
\end{equation}%
where

\begin{equation}
\varepsilon _{AB}=\left( 
\begin{array}{cc}
0 & 1 \\ 
-1 & 0%
\end{array}%
\right) .  \label{51}
\end{equation}

The identification of surreal numbers with rebits means that its
complexification must be related to qubit theory and therefore with twistor
theory. So one has discovered that the use of the completely antisymmetric
object epsilon $\varepsilon ^{a_{1}...a_{d}}$ allows to define the Plucker
coordinates which must to satisfy the Grassmann-Pl\"{u}cker relation. In
turn, we have proved that this relation is a common mathematical central
notion in oriented matroids, qubit theoy, twistor theory and surreal number
theory.

Moreover, it has been proved in Refs. [27]-[29] that for normalized qubits
the complex $1$-qubit, $2$-qubit and $3$-qubit are deeply related to
division algebras via the Hopf maps, $S^{3}\overset{S^{1}}{\longrightarrow }%
S^{2}$, $S^{7}\overset{S^{3}}{\longrightarrow }S^{4}$ and $S^{15}\overset{%
S^{7}}{\longrightarrow }S^{8}$, respectively. It seems that there does not
exist a Hopf map for higher $N$-qubit states. So, from the perspective of
Hopf maps, and therefore of division algebras, one arrives to the conclusion
that $1$-qubit, $2$-qubit and $3$-qubit are more special than higher
dimensional qubits (see Refs. [27]-[29] for details). Again one wonders
whether surreal numbers can contribute in this qubits theory framework.

The original idea of Penrose was to replace the continuity of the Minkowski
space for new geometric framework which may allow for a discrete structure
and in this way unify general relativity and quantum mechanics. In fact, one
of the original motivation to introduce twistors was be able to have
mathematical arena in which the discretization of the spacetime was
possible. The hope was that the complex structure of twistors may be
connected with quantum mechanics. In a sense the idea was to replace $R^{4}$
by $C^{4}$ and in this way, since the object in $C^{4}$ are complex, one may
be able to connect with quantum mechanics which intrinsically is a complex
structure. Ironically, according to the discussion in this work, it seems to
us that the combinatorial structure searched by Penrose in connection with
quantum gravity is not the twistors itself but the underlying oriented
matroid theory. But ground set in oriented matroids can be constructed by
strings of the set $\{+-\}$ which are the main tool in qubit theory and
surreal numbers. All these comments suggested that the concepts such as
chirotopes (phirotopes), qubits, twistors and surreals must be considered
mathematical tools underlying quantum gravity.

Let us analysis deeply the connection between surreal numbers and qubits.
For this purpose we shall assume that one may be able to write a surreal
complex numbers $\mathcal{Z}$ in the form

\begin{equation}
\mathcal{Z}=\mathcal{J}_{1}+i\mathcal{J}_{2},  \label{52}
\end{equation}%
where $\mathcal{J}_{1}$ and $\mathcal{J}_{2}$ are two surreal numbers
according to the formula (43). This complexification of surreal numbers must
establish a complete connection with the $N$-qubit structure if one assume
the existence of a complex operator $\mathcal{\hat{Z}}_{A_{1}A_{2}...A_{N}}$
such that%
\begin{equation}
\mathcal{\hat{Z}}_{A_{1}A_{2}...A_{N}}\mid
A_{1}A_{2}...A_{N}>=\dsum%
\limits_{A_{1},A_{2},...,A_{N}=0}^{1}Q_{A_{1}A_{2}...A_{N}}\mid
A_{1}A_{2}...A_{N}>=\mathcal{J}\mid A_{1}A_{2}...A_{N}>.  \label{53}
\end{equation}%
This is inspired in the observation that $\mathcal{J}$ in (43) can be
associated with the eigenvalues of a $z$-component $\hat{J}_{z}$ of the
total angular momentum $\hat{J}$ in quantum mechanics. Of course in such
case one has $J_{z}=l\pm \frac{1}{2}$, with the identification of $\frac{1}{2%
}$-spin of the system. The surprise with surreal numbers is that predicts
that besides $\frac{1}{2}$-spin system there must exist infinite number of $%
\mathcal{J}$-spins, according to the formula (43). Thus, for instance one
must include particles with $\frac{1}{4}$-spin (see Refs. [30] and [31]) and 
$\frac{1}{8}$-spin and in general particles with dyadic rational $\frac{m}{%
2^{n}}$-spin.

Traditionally, quantum mechanics enter in the above twistor formalism when
one writes all possible gauge fields (and their associated field equations)
in twistor language and proceed to quantize in the usual way. In the case of
qubit theory things are different because, even from the begining, qubits
refers to quantum states. Thus, concepts of quantum mechanics such as the
density of states are constructed from the corresponding entanglement
monotones [23]. Here, we would like to propose an alternative possible route
to connect further our formalism with quantum mechanics. The central idea is
to continue looking the surreal numbers as a quantities associated with
different dyadic spins ($\frac{m}{2^{n}}$-spin). Let us explain in some
detail this idea. As we mentioned $\mathcal{J}$ in (43) seems to play the
analogue of the eigenvalues of the $z$-component $\hat{J}_{z}$ of the
angular momentum operator, namely $J_{z}=l\pm \frac{1}{2}$. Roughly
speaking, from the point of view of number theory, the quantization of a
physical system means to go from the real numbers (continuum) $R$ to natural
numbers $N$ (discrete). In the case of surreal numbers things are different
because one starts with the $0$-day, $1$-day, $2$-day and so on and in the $%
\omega $-day (this is the way mathematitians called) one obtains the real
numbers $R$. In other words one starts with a discrete structure and finds
the continuum scenario. Moreover, if in addition to (43) one uses the
identity

\begin{equation}
2^{n+1}=2+2+4+8+...+2^{n},  \label{54}
\end{equation}%
it is not difficult to show that $\mathcal{J}$ in (43) satisfies the
expression

\begin{equation}
-l<\mathcal{J}<l.  \label{55}
\end{equation}%
Since $l<j$ one also has

\begin{equation}
-j<\mathcal{J}<j.  \label{56}
\end{equation}%
Here, one assumes that from (43) one has $j=l+s$. Of course, (56) is the
analoguos inequality of the eigenvalue of the total angular momentum.
Following this route of thoughts one first note that surreal numbers of the
type $(++...++)$ (or the corresponding negative part) can be associated with
higher integer-spins, $1,2,3,...$, while surreal numbers of the type $%
(++...+-)$ can be associated with half-inter spins, $1/2,3/2,5/2,...$. This
means that in principle bosons and fermions are part of the surreal
structure and therefore supersymmetry must be present. Thus one must expect
that a generalized supersymmetry can be obtained if one includes other
surreal numbers such as $1/4,3/4,1/8,3/8,$ and so on. Since, as we
mentioned, the dyadic rational $m/2^{n}$ are dense in the reals $R$ one
should expect that eventually, in the $\omega $-day, the anyons may emerge.
What about the graviton? This corresponds to the surreal number $2$ or $2$%
-spin. Thus, just as in string theory the graviton is just one resonance out
of many or even infinity resonances, in our case the graviton is just a
physical system with particular value $2$-spin, but in principle one has all
kind of dyadic-spin particles. Thus, according to these observations it
seems that quantum gravity should not be seen as an isolated problem but as
part of a much larger system in which all types of dyadic-spins are present.

Another source of interesting developments it may emerge from the analysis
of singularities, both in balck-holes and cosmology. In fact, from the point
of view of surreal numbers theory the black-hole singularity $%
2MG/c^{2}r\rightarrow \infty $, when $r\rightarrow 0$, and the Big-Bang
singularity (of the radiation energy density) $\rho _{r}=\rho
_{0}/a^{4}\rightarrow \infty $, when $a\rightarrow 0$ are not a real problem
because in such a mathematical theory all kind of infinite large and
infinite small are present.

It is worth mentioning that in the Ref. [32] the twistor space and the Pl%
\"{u}cker coordinates are used to determine the geometry of the instantons
solutions of Yang-Mills theory. It may interesting for further research to
find the connection between instantons formalism and surreal number theory.

Finally, let us just mention that using fiber bundle concept in oriented
matroid theory [33] and [34] a connection with $p$-branes and phirotopes was
established [6]. Thus according to the present development one may expect
that eventually a link between $p$-branes and surreal numbers must be route
to follow in the quest of quantum gravity.

\bigskip \ 

\noindent \textbf{Acknowledgments: }I would like to thank the Mathematical,
Computational \& Modeling Sciences Center of the Arizona State University
where part of this work was developed. I would like also to thank the two
referees and the editor for valuable comments.

\smallskip \

\end{document}